\documentclass[conference]{IEEEtran}

\usepackage{graphicx}
\usepackage{amssymb}
\usepackage{amsmath}
\usepackage{epsfig}
\usepackage{cite}
\usepackage{verbatim}
\usepackage{enumerate}
\usepackage{subfigure}
\usepackage{multicol}
\usepackage{pstricks}
\usepackage{pst-node}

\newcommand{\be}{\begin{equation}}
\newcommand{\ee}{\end{equation}}

\newtheorem{theorem}{\bf Theorem}

\newtheorem{corollary}{\bf Corollary}

\begin{document}
\title{Degrees of Freedom in Multiuser MIMO}
\author{\authorblockN{Syed A. Jafar}
\authorblockA{Electrical Engineering and Computer Science\\
University of California Irvine, California, 92697-2625\\
Email: syed@ece.uci.edu\\ \vspace{-1cm}}
\and
\authorblockN{Maralle J. Fakhereddin}
\authorblockA{Department of Electrical Engineering \\
California Institute of Technology, Pasadena, CA 91125 \\
Email: maralle@systems.caltech.edu\\ \vspace{-1cm}
}}

\maketitle
\begin{abstract}
We explore the available degrees of freedom for various multiuser MIMO communication scenarios such as the multiple access, broadcast, interference, relay, $X$ and $Z$ channels. For the two user MIMO interference channel, we find a general inner bound and a genie-aided outer bound that give us the exact number of degrees of freedom in many cases. We also study a share-and-transmit scheme for transmitter cooperation. For the share-and-transmit scheme, we show how the gains of transmitter cooperation are entirely offset by the cost of enabling that cooperation so that the available degrees of freedom are not increased.
\end{abstract}
\section{Introduction}\label{section:introduction} 
Multiple input multiple output (MIMO) systems have assumed great importance in recent times because of their remarkably higher capacity compared to single input single output (SISO) systems. It is well known \cite{Foschini_Gans, Telatar,Zheng_Tse} that the capacity of a point-to-point MIMO system with $M$ inputs and $N$ outputs increases linearly as $\min(M,N)$ at high SNR. For power and bandwidth limited wireless systems this opens up another dimension - ``space'' that can be exploited in a similar way as time and frequency. Similar to time division and frequency division multiplexing, MIMO systems present the possibility of multiplexing signals in space. For example, using the singular value decomposition of a MIMO channel, one can generate parallel channels in space similar to the multiple channels created by dividing time or frequency into orthogonal slots. 

Inspite of the theoretical duality of time, frequency, and space, the contextual constraints of a communication system create important differences that determine whether the degrees of freedom that exist in each of these dimensions are \emph{available} or not. In particular, the availability of the spatial degrees of freedom depends upon two factors: cooperation within inputs and outputs, and channel knowledge. Previous work has shown that in the absence of channel knowledge the spatial degrees of freedom are lost \cite{Syed_IT, Amos_IT}. In this paper we will focus on  communication scenarios with constrained cooperation between inputs and outputs distributed among multiple users. What makes the multiuser MIMO systems especially challenging is that unlike the point-to-point case, joint processing is not possible at the inputs or the outputs controlled by different users. The available spatial degrees of freedom are affected by the inability to jointly process the signals at the distributed inputs and outputs. 

In this paper we quantify the loss in the available degrees of freedom under the distributed processing constraints imposed by various multiuser communication scenarios.  It was recently shown in \cite{Madsen_CTW} that cooperation between single antenna transmitters  does not provide additional multiplexing gain in an interference channel. In this paper, we explore the benefits of cooperation between the transmitters when the nodes have \emph{multiple} antennas. We establish a general innerbound and a genie-based outerbound on the number of degrees of freedom for a multiple antenna interference channel. For many cases of practical interest, these bounds are shown to be tight and we have the exact number of degrees of freedom. We also consider a simple cooperative scheme to understand why cooperation between transmit nodes does not increase the degrees of freedom. Through this simple scheme we are able to show how the benefits of cooperation are completely offset by the cost of enabling the cooperation. The extensions of the results to the X-channel, the Z-channel and the relay channel are also discussed.

\section{Degrees of Freedom Measure}\label{section:DOF}
In order to isolate the impact of distributed processing from the channel uncertainty we assume that the channel state is fixed and perfectly known at all transmitters and receivers. Also, we assume that the channel matrices are sampled from a rich scattering environment. Therefore we can ignore the measure zero event that some channel matrices are rank deficient. It is well known that the capacity of a \emph{scalar} AWGN channel scales as $\log(SNR)$ at high SNR. On the other hand, for a single user MIMO channel with $M$ inputs and $N$ outputs, the capacity growth rate can be shown to be $\min(M,N)\log(SNR)$ at high SNR. This motivates the natural definition of the spatial degrees of freedom as:
\begin{equation}
\eta\triangleq \lim_{\rho\rightarrow\infty}\frac{C_\Sigma(\rho)}{\log(\rho)},
\end{equation}
where $C_\Sigma(\rho)$ is the sum capacity (just the capacity in case of point to point (PTP) communications) at SNR $\rho$. In other words, the degrees of freedom $\eta$ represent the maximum \emph{multiplexing gain} \cite{Zheng_Tse} of the generalized MIMO system. For the point to point case, the $(M,N)$ degrees of freedom are easily seen to correspond to the parallel channels that can be isolated using the SVD operation, involving joint processing at the $M$ inputs and joint processing at the $N$ outputs, i.e.,
\begin{eqnarray}
\eta(\mbox{PTP})&=&\min(M,N)
\end{eqnarray}
\section{The Multiple Access Channel}\label{section:MAC}
The multiple access channel is an example of a MIMO system where cooperation is allowed only between the channel outputs. Let the MAC consist of $N$ outputs controlled by the same receiver and $2$ users, each controlling $M_1$ and $M_2$ inputs for a total of $M=M_1+M_2$ inputs. For the MAC the available degrees of freedom are the same as with perfect cooperation between all the users.
\begin{equation}
\eta(\mbox{MAC})=\eta(\mbox{PTP})=\min(M_1+M_2,N).
\end{equation}
While the capacity region of the MIMO MAC is well known and the spatial multiplexing gain has also been explored in previous work, we include the following constructive proof to introduce zero forcing notation which will be useful in the interference channel. Zero forcing, which is normally a suboptimal strategy, is sufficient in this case (as well as on the MIMO broadcast channel)  to utilize all the degrees of freedom. 

\noindent{\it Converse}: The converse is straightforward because, for the 
same number of inputs and outputs, $\eta(\mbox{MAC})\leq \eta(\mbox{PTP})=\min(M_1+M_2,N)$. In other words, the lack of cooperation at the inputs can not increase the degrees of freedom.\\
{\it Achievability}: The $N\times 1$ received signal ${\bf Y}$ at the MAC receiver
\begin{equation}
{\bf Y}=\sum_{k=1}^2{\bf H}^{(k)}{\bf X}^{(k)}+{\bf N} = {\bf V_H}^\dagger{\bf V_X}+{\bf Z},
\end{equation}
where ${\bf N}$ is the $N\times 1$ additive white Gaussian noise (AWGN) vector, ${\bf H}^{(k)}$ is the $N\times M_k$ channel matrix for user $k$, and ${\bf X}^{(k)}$ is the $M_k\times 1$ transmitted vector for user $k$. ${\bf V_H} = V({\bf H}^{(\cdot)^\dagger})$ is the $(M_1+M_2)\times N$ matrix obtained by vertically stacking the matrices ${\bf H}^{(1)^\dagger}$ and ${\bf H}^{(2)^\dagger}$. Similarly, ${\bf V_X}=V({\bf X}^{(\cdot)})$ is the $(M_1+M_2)\times 1$ matrix obtained by vertically stacking ${\bf X}^{(1)}$ and ${\bf X}^{(2)}$. Transforming the output vector 
\[
{\bf Y}^{\mbox{new}}= \left({\bf V_H V_H}^\dagger\right)^{-1}{\bf V_H}{\bf Y}\]
(using generalized Moore-Penrose inverse) and ignoring the zero gain channels results in the $\min(M,N)$ parallel channels
\begin{equation}
{\bf Y}^{\mbox{new}}(i)={\bf V_X}(i)+{\bf N}^{\mbox{new}}(i), ~~~~~1\leq i\leq\min(M,N),
\end{equation}
where ${\bf N}^{\mbox{new}}(i)\sim\mathcal{N}(0,\lambda_i)$ are Gaussian noise terms and $\lambda_i$ is the $i^{th}$ diagonal term of $\left({\bf V_H V_H}^\dagger\right)^{-1}$. The noise terms may be correlated across different channels but the correlations are inconsequential since each channel is encoded and decoded separately. Dividing power equally among the $\min(M,N)$ channels, we can achieve
\begin{equation}
\eta(\mbox{MAC})\geq \lim_{\rho\rightarrow\infty}\frac{1}{\log(\rho)}\sum_{i=1}^{\min(M,N)}\log\left(1+\frac{\rho}{\min(M,N)}\frac{1}{\lambda_i^2}\right)\nonumber
\end{equation}
\begin{eqnarray}
~~~&=&\lim_{\rho\rightarrow\infty}\frac{1}{\log(\rho)}\left[\min(M,N)\log(\rho)+
\right.\nonumber\\
&&\left.\sum_{i=1}^{\min(M,N)}\log\left(\frac{1}{\lambda_i^2\min(M,N)}\right)\right]=\min(M,N)\nonumber
\end{eqnarray}
Note that the channel gains or the exact power allocation does not affect the degrees of freedom as long as the SNR on each channel is proportional to $\rho$.

Combining the converse and the achievability, we have established that  $\eta({MAC})=\min(M_1+M_2,N)$.
\section{The Broadcast Channel}\label{section:BC}
The broadcast channel is an example of a MIMO system where cooperation is allowed only between the channel inputs. Let the BC consist of $M$ inputs controlled by the same transmitter and $2$ users, each controlling $N_1$ and $N_2$ outputs for a total of $N=N_1+N_2$ outputs. Similar to the MAC, it is possible to show that by zero forcing at the BC transmitter, $\min(M,N)$ parallel channels can be created, so that the total degrees of freedom are the same as with perfect cooperation between all the users.
\begin{equation}
\eta(\mbox{BC})=\eta(\mbox{MAC})=\eta(\mbox{PTP})=\min(M,N).
\end{equation}

\section{Interference Channel}\label{section:INT}
Consider an $(M_1,N_1),(M_2,N_2)$ interference channel with two transmitters $T_1$ and $T_2$, and two receivers $R_1$ and $R_2$, where $T_1$ has a message for $R_1$ only and $T_2$ has a message for $R_2$ only. $T_1$ and $T_2$ have $M_1$ and $M_2$ antennas respectively. $R_1$ and $R_2$ have $N_1$ and $N_2$ antennas respectively. Assume that we arrange the links so that link 1 between $T_1$ and $R_1$ always has the most number of antennas either at its transmitter or receiver. We denote the channel for link 1 with the $N_1$x$M_1$ channel gain matrix $\rm{H}^{(1)}$. Similarly, the channel for link 2 is denoted by $N_2$x$M_2$ channel gain $\rm{H}^{(2)}$. The channel between $T_1$ and $R_2$ is described by $N_2$x$M_1$ channel gain matrix $\rm{Z}^{(2)}$, and the channel between $T_2$ and $R_1$ by $N_1$x$M_2$ channel gain matrix $\rm{Z}^{(1)}$. Figure 1 shows an illustration of this interference channel. 

\begin{figure}
\begin{center}
\includegraphics[height=6cm,width=2in]{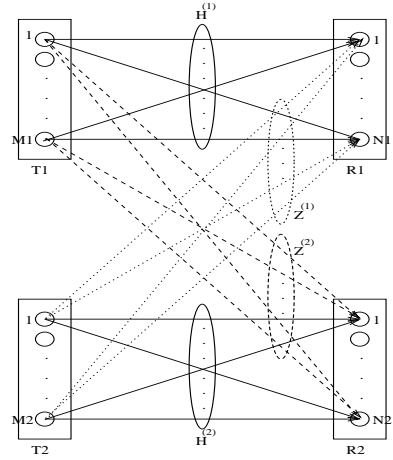}
\end{center}
\caption{Interference channel with ($M_1$,$N_1$) and ($M_2$,$N_2$)}
\end{figure}

\subsection{Innerbound on the Available Degrees of Freedom}
For the $(M_1,N_1),(M_2,N_2)$ interference channel we prove the following innerbound on the available degrees of freedom.
\begin{eqnarray}\label{eqnarray:inner}
\eta(\mbox{INT}) &\geq& \min(M_1,N_1) \nonumber\\ 
    &+& \min(M_2-N_1,N_2)^{+}~1(M_1>N_1)  \nonumber\\
    &+& \min(M_2,N_2-M_1)^{+}~1(M_1<N_1),
\end{eqnarray}
where 1(.) is the indicator function and $(x)^+=\max(0,x)$. While we conjecture that this bound is tight for any $M_1,N_1,M_2,N_2$ we can prove a converse only with some additional assumptions on the number of antennas. A general achievability proof is outlined next.

\subsubsection{Sketch of Achievability Proof}
According to our system model, either $M_1\geq N_1,M_2,N_2$ or $N_1>M_1,M_2,N_2$. First, we consider the case when $M_1 \geq N_1,M_2,N_2$.

\noindent{\it Step 1:} From SVD, $Z^{(2)}$ can be described as $Z^{(2)} = U \Lambda V^{H}$, where U and V are $N_2$x$N_2$ and $M_1$x$M_1$ unitary matrices respectively and $\Lambda$ is the diagonal matrix of singular values of $Z^{(2)}$. By applying SVD to $Z^{(2)}$, we decompose the channel into $min(M_1,N_2)$ parallel channels between $T_1$ and $R_2$. Following SVD, there are $M_1-N_2$ effective inputs at $T_1$ that are not connected to $R_2$, and therefore do not cause any interference to $R_2$. 

\noindent{\it Step 2:} Similarly, applying SVD to the ($M_2$,$N_1$) channel between $T_2$ and $R_1$, $Z^{(1)}$, creates $min(M_2,N_1)$ parallel connections. There are $(M_2-N_1)^{+}$ effective inputs at $T_2$ that are not connected to $R_1$, and therefore do not cause any interference with $R_1$. 

\noindent{\it Step 3:} For link 1's communication between $T_1$ and $R_1$, all $N_1$ effective outputs are used by $R_1$. 

\noindent{\it Step 4:} $T_1$ transmits to $R_1$ using $N_1$ effective inputs such that at most $(N_1+N_2-M_1)^{+}$ effective inputs that are active are also connected to $R_2$.
\noindent{\it Step 5:} $T_2$ and $R_2$ use only those effective inputs or outputs that are not connected to an active effective input or output of link 1. 

\noindent{\it Step 6:} Link 1 is left with $N_1$ effective inputs and $N_1$ effective outputs, i.e. the number of degrees of freedom for link 1 $= N_1$.

\noindent{\it Step 7:} For link 2, $T_2$ is left with $(M_2-N_1)^{+}$ effective inputs while $R_2$ is left with $min(M_1-N_1,N_2)$ effective outputs, i.e. the number of degrees of freedom for link 2 $= min(M_2-N_1,min(M_1-N_1,N_2))^{+} = min(M_2-N_1,N_2)^{+}$ since $M_1 \geq M_2$ by assumption. Hence proved. 

\begin{figure}
\begin{center}
\includegraphics[height=6.0cm,width=\linewidth]{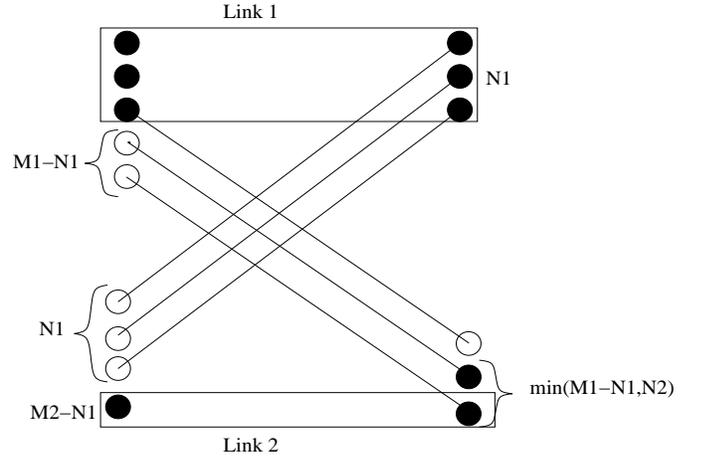}
\end{center}
\caption{Achievability proof for Interference channel with ($M_1$,$N_1$) and ($M_2$,$N_2$) when $M_1 \geq M_2,N_1,N_2$}
\end{figure}

For the case when $N_1>M_1,M_2,N_2$, the same logic is followed. After SVD, step 1 results in $min(M_1,N_2)$ parallel connections. Similarly, in step 2 SVD creates $min(M_2,N_1)$ such connections. All $M_1$ effective inputs are used for communication on link 1. There are $N_1-M_1$ effective outputs left at $R_1$ such that at most $(M_1+M_2-N_1)^{+}$ effective outputs that are active and also connected to $T_1$. $T_2$ has $min(M_2,N_1-M_1)$ effective inputs and $R_2$ has $(N_2-M_1)^{+}$ effective outputs for communication on link 2. Therefore, the number of degrees of freedom for link 2 $= min(min(M_2,N_1-M_1),N_2-M_1)^{+}$. Since by assumption $N_1 \geq N_2$, the total number of degrees of freedom for this case is $min(M_1,N_1)+min(M_2,N_2-M_1)^{+}$.

By adding the results from the above cases, we obtain a general achievable proof of (\ref{eqnarray:inner}). An illustration of this proof is shown in figure 2.

\subsection{Outerbounds on the Available Degrees of Freedom}
To start with, notice that a trivial outerbound is obtained from the point to point case, i.e. $\eta(\mbox{INT})\leq \min(M_1+M_2,N_1+N_2)$. Indeed this outerbound coincides with the innerbound when either $\min(M_1,M_2)\geq N_1+N_2$ or $\min(N_1,N_2)\geq M_1+M_2$.

In general, while the capacity region of the interference channel is not known  even with single antennas at all the nodes, various outerbounds have been obtained \cite{Carliel_outer,Kramer_Outer,Jafar_Vishwanath_ITW} that have been useful in finding the capacity region in some special cases \cite{Ahlswede_twin,Carliel_int}. Most of the existing outerbounds are for single antenna systems. 

For our purpose we develop a genie based outerbound for a MIMO interference channel where the number of antennas at either receiver is greater than or equal to the number of transmit antennas at the interfering transmitter. In other words, we develop an outerbound for the case when either $N_1\geq M_2$ or $N_2\geq M_1$. We find that, in many cases, this outerbound is sufficiently tight  to establish the number of degrees of freedom.

For this section, since we do not use the assumption that $\max(M_1,N_1)\geq\max(M_2,N_2)$, the proof for the cases $N_1\geq M_2$ or $N_2\geq M_1$ are identical. Therefore, we state the following theorem under the assumption $N_1\geq M_2$.
\begin{theorem}\label{theorem:outer}
For the $(M_1,N_1,M_2,N_2)$ interference channel with $N_1\geq M_2$, the sum capacity is bounded above by the sum capacity of the corresponding $(M_1, M_2, N_1)$ multiple access channel with additive noise ${\bf N^{(1)}}\sim\mathcal{N}({\bf 0, I_N})$ modified to ${\bf N^{(1)'}}\sim\mathcal{N}({\bf 0, K^{'}})$ where
\begin{eqnarray*}
{\bf K^{'}}&=& {\bf I_N}- {\bf Z^{(1)}}\left({\bf Z^{(1)\dagger}}{\bf Z^{(1)}}\right)^{-1}{\bf Z^{(1)\dagger}}+\alpha{\bf Z^{(1)}}{\bf Z^{(1)^\dagger}},\\
\alpha & = & \min\left(\frac{1}{\sigma^2_{\max}({\bf Z^{(1)}})},\frac{1}{\sigma^2_{\max}({\bf H^{(2)}})}\right).
\end{eqnarray*}
\end{theorem}
\proof

Let us define
\begin{eqnarray*}
{\bf N^{(1)}_a}&\sim&\mathcal{N}\left({\bf 0}, {\bf I_N}- {\bf Z^{(1)}}\left({\bf Z^{(1)\dagger}}{\bf Z^{(1)}}\right)^{-1}{\bf Z^{(1)\dagger}}\right)\\
{\bf N^{(1)}_b}&\sim&\mathcal{N}\left({\bf 0}, {\bf Z^{(1)}}\left({\bf Z^{(1)\dagger}}{\bf Z^{(1)}}\right)^{-1}{\bf Z^{(1)\dagger}}-\alpha{\bf Z^{(1)}}{\bf Z^{(1)^\dagger}}\right)\\
{\bf N^{(1)}_c}&\sim&\mathcal{N}\left({\bf 0},\alpha{\bf Z^{(1)}}{\bf Z^{(1)^\dagger}}\right),
\end{eqnarray*}
as three $N \times 1$ jointly Gaussian and mutually independent random vectors. The positive semidefinite property of the respective covariance matrices is easily established from the definition of $\alpha$. 

Without loss of generality we assume 
\begin{eqnarray*}
{\bf N^{(1)}}&=&{\bf N^{(1)}_a}+{\bf N^{(1)}_b}+{\bf N^{(1)}_c}\\
{\bf N^{(1)'}}&=&{\bf N^{(1)}_a}+{\bf N^{(1)}_c}
\end{eqnarray*}
Furthermore, because ${\bf N^{(1)}}$ and ${\bf N^{(2)}}$ have the same marginal distributions and the capacity of the interference channel does not depend on the correlation between ${\bf N^{(1)}}$ and ${\bf N^{(2)}}$, the capacity region is not affected if we assume
\begin{eqnarray*}
{\bf N^{(1)}}&=&{\bf N^{(2)}}.
\end{eqnarray*}

Since a part of the proof is similar to the corresponding proof for the single antenna case we will summarize the common steps, and emphasize only the part that is unique to MIMO interference channels. Consider any achievable scheme for any rate point within the capacity region of the interference channel, so that $R_1$ and $R_2$ can correctly decode their intended messages from their received signals with sufficiently high probability. 

\noindent{\it Step 1:} We replace the original additive noise ${\bf N^{(1)}}$ at $R_1$ with ${\bf N^{(1)'}}$ as defined in Theorem \ref{theorem:outer}. We argue that this does not make the capacity region smaller because the original noise statistics can easily be obtained by locally generating and adding noise ${\bf N^{(1)}_b}$ at $R_1$. Therefore, since $R_1$ was originally capable of decoding its intended message with noise ${\bf N^{(1)}}$, it is still capable of decoding its intended message with noise ${\bf N^{(1)'}}$.

\noindent{\it Step 2:} Suppose that a genie provides $R_2$ with side information containing the entire codeword ${\bf X}^{(1)}$. Since ${\bf X}^{(2)}$ is independent of ${\bf X}^{(1)}$ receiver  $R_2$ simply subtracts out the interference from its received signal. Thus, the channel ${\bf Z}^{(2)}$ can be eliminated without making the capacity region smaller.

\noindent{\it Step 3:} By our assumption, $R_1$ can decode its own message and therefore it can subtract ${\bf X}^{(1)}$ from its own received signal as well. In this manner, after the interfering signals have been subtracted out we have
\begin{eqnarray*}
{\bf Y}^{(1)}&=&{\bf Z^{(1)}}{\bf X^{(2)}}+{\bf N^{(1)'}},\\
{\bf Y}^{(2)}&=&{\bf H^{(2)}}{\bf X^{(2)}}+{\bf N^{(2)}}.
\end{eqnarray*}
To complete the proof we need to show that if $R_2$ can decode ${\bf X}^{(2)}$ then so can $R_1$. This would imply that $R_1$ can decode both messages, hence giving us the MAC outer bound. 

\noindent{\it Step 4:} Without loss of generality let us perform a singular value decomposition ${\bf H}^{(2)}={\bf U}^{(2)}{\bf \Lambda^{(2)}}{\bf V^{(2)}}$ on the channel  between $T_2$ and $R_2$. This is a lossless operation that leads to:
\begin{equation}
{\bf Y}^{(2)\mbox{new}}={\bf X^{(2)\mbox{new}}}+\left({\bf \Lambda^{(2)}}\right)^{-1}{\bf N^{(2)}},
\end{equation}
where ${\bf X^{(2)\mbox{new}}}={\bf V^{(2)}}{\bf X^{(2)}}$.

To save space we allow some notation abuse as we use generalized inverse and ignore the terms that correspond to zero diagonal channel gains in ${\bf \Lambda^{(2)}}$. Note that these channels are useless for $R_2$. Also, we use the same symbol for rotated versions of noise that are statistically equivalent.

\noindent{\it Step 5:} Next, we show that $R_1$ can obtain a stronger channel to ${\bf X^{(2)\mbox{new}}}$ so that if $R_2$ can decode it, so can $R_1$. To this end, let $R_1$ use zero forcing to obtain:
\begin{eqnarray*}
{\bf Y}^{(1)\mbox{new}}&=&{\bf X^{(2)\mbox{new}}}+{\bf V^{(2)}}\left({\bf Z^{(1)\dagger}}{\bf Z^{(1)}}\right)^{-1}{\bf Z^{(1)\dagger}}{\bf N^{(1)'}},\\
&=&{\bf X^{(2)\mbox{new}}}+\alpha{\bf N^{(2)}}
\end{eqnarray*}

Now both $R_1$ and $R_2$ have a diagonal channel with input ${\bf X^{(2)\mbox{new}}}$ and uncorrelated additive white noise components on each diagonal channel. Moreover, the strongest channel for $R_2$ has noise $\frac{1}{\sigma^2_{\max}({\bf H^{(2)}})}$. However the noise on any channel for $R_1$ is only $\alpha$ which is smaller. Thus, we argue once again that $R_1$ can locally generate noise and add it to its received signal to create a statistically equivalent noise signal as seen by $R_2$. In other words, $R_1$ has a less noisy channel to $T_2$ and therefore can decode any signal that $R_2$ can. Since $R_1$ can decode $T_1$'s message by assumption, we have the MAC outerbound. \hfill\QED

The MAC outerbound leads directly to the following outerbound on the number of degrees of freedom.
\begin{corollary}
For the $(M_1,N_1,M_2,N_2)$ interference channel with $N_1\geq M_2$, the number of degrees of freedom $\eta(\mbox{INT})\leq \min(M_1+M_2, N_1)$. Similarly, if $N_2\geq M_1$, then $\eta(\mbox{INT})\leq \min(M_1+M_2, N_2)$.
\end{corollary}
The outerbound and the innerbound are tight in many cases where we have the exact number of degrees of freedom. Some examples are provided in the following table.
\begin{equation*}
\begin{array}{|c|c|c|}
\hline
(M_1,N_1) & (M_2,N_2)& \eta(INT) \\
\hline
(1, 1) & (1, 1) & 1 \\
\hline
(1,2) & (1, 2) & 2\\
\hline
(2,1) & (2,1) & 2\\
\hline
(1,2) & (2,1) & 1\\
\hline
(3,2) & (2,3) & 2\\
\hline
(2,3) & (2,3) & 3\\
\hline
(2,3) & (1,3) & 3\\
\hline
(2,2) & (3,2) & 2\\
\hline
\end{array}
\end{equation*}
\section{Effect of transmit cooperation on the number of degrees of freedom}\label{section:COOP}
Comparing the interference channel and the broadcast channel obtained by full cooperation between the transmitters it is clear that the available degrees of freedom are severely limited by the lack of transmitter cooperation in the interference channel. As an example, consider the interference channel with $(M_1,N_1)=(n,1)$ and $(M_2,N_2)=(1,n)$. From the preceding section we know there is only one available degree of freedom in this channel. However, if full cooperation between the transmitters is possible the resulting broadcast channel has $(M, N_1,N_2)=(n+1, 1, n)$. The number of degrees of freedom is now $n+1$. Therefore, transmitter cooperation would seem highly desirable. Rather surprisingly, it has been shown recently \cite{Madsen_CTW} that for the $(1,1,1,1)$ interference channel, allowing the transmitters to cooperate through a wireless link between them (even with full duplex operation), does not increase the degrees of freedom. For MIMO interference channels, as suggested by the example above, the potential benefits of cooperation are even stronger and it is not known if transmitter cooperation can increase the degrees of freedom. The capacity results of \cite{Madsen_CTW}  do not seem to allow direct extensions to MIMO interference channels. 

To gain insights into the cost and benefits of cooperation in a MIMO interference channel, we consider a specific scheme where the transmitters first share their information in a full duplex mode as a MIMO channel (step 1) and subsequently transmit together as a broadcast channel to the receivers. We will refer to this scheme as the share-and-transmit scheme. 


\subsection{Degrees of Freedom with Share-and-Transmit}\label{subsection:SNT}
Consider an interference channel with two transmitters and two receivers. Suppose each transmitter is equiped with $M$ antennas and the receivers have $N$ antennas each ($M\leq N$). Also assume that each transmitter is sending information with rate $R$. Note that while we make the preceding simplifying assumptions for simplicity of exposition, the following analysis and the main result extend directly to the general case of unequal number of antennas and unequal rates.

From (\ref{eqnarray:inner}), we know that the number of degrees of freedom for this interefernce channel with no transmitter cooperation is $\min(M,N)+\min(M,N-M)^{+} = M+\min(M,N-M)^{+}$. For the share-and-transmit scheme, we compute the degrees of freedom as follows. We first find the capacity of the sharing link $C_s$ and the capacity of transmission $C_t$. Then, we find the total capacity of the system $C$ by evaluating the total amount of data transmitted divided by the total time it takes to transmit this data. In other words,
\begin{equation}
C=\frac{2R}{\frac{R}{C_s}+\frac{2R}{C_t}}~~.\ee
Dividing by log(SNR) where SNR is large, we obtain the total number of degrees of freedom as
\be\label{DOF_SNT}
\underset{SNR \rightarrow \infty}{\lim}~\frac{C}{\log{SNR}}=\frac{2}{\frac{1}{DOF(sharing)}+\frac{2}{DOF(transmit)}}~.\ee The number of degrees of freedom for the sharing link is that of a multiple antenna point-to-point channel with $M$ transmit and $M$ receive antennas. This is well known to be $min(M,M) = M$. After the transmitters share their information, they can fully cooperate in sending their information in step 2. As a result, we can consider the channel as equivalent to a broadcast channel with one transmitter with $2M$ antennas and two users each with $N$ antennas. The number of degrees of freedom for this channel is therefore $\min(2M,2N) = 2\min(M,N)$. Therefore (\ref{DOF_SNT}), which gives the total number of degrees of freedom for the share-and-transmit scheme, becomes $\frac{2M \min(M,N)}{M + \min(M,N)} = M$. This is always less or equal to  the number of degrees of freedom given from (\ref{eqnarray:inner}) for the original (transmit only) scheme. In other words,
\be
M+\min(M,N-M)^{+} \geq M.\ee  Therefore, we conclude that (under the system constraints described in this section) transmit cooperation in the high SNR regime does not provide any advantage to the number of degrees of freedom in the multiple antenna interference channel. In the following section, we verify this result by simulations, and discuss the effect of transmit cooperation when the sharing links between the transmitters are stronger than their transmission links to the receivers.


\section{Simulation Results}\label{section:sim}
\begin{figure}
\begin{center}
\includegraphics[height=5.0cm,width=\linewidth]{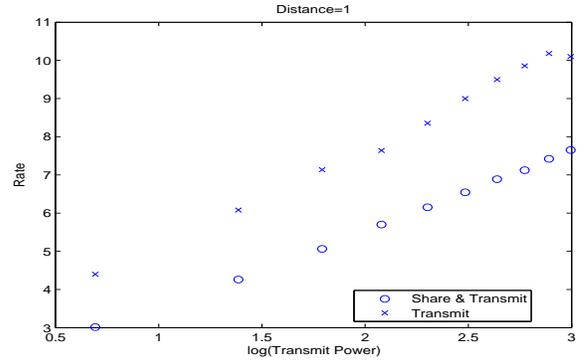}
\end{center}
\caption{Rate versus log(Transmit Power) for (4,1)(4,1) Interference channel with Distance=1 between transmitters and between each transmitter and receiver.}
\end{figure}

\begin{figure}
\begin{center}
\includegraphics[height=5.0cm,width=\linewidth]{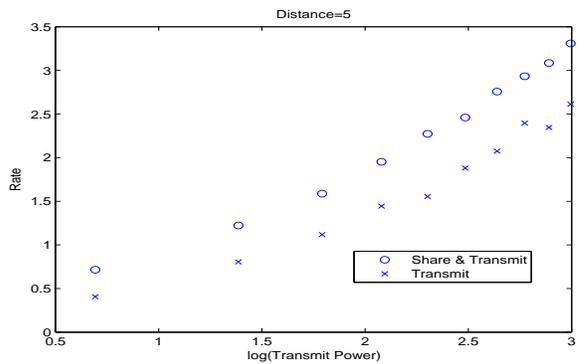}
\end{center}
\caption{Rate versus log(Transmit Power) for (4,1)(4,1) Interference channel with Distance=1 between transmitters and Distance=5 between each transmitter and receiver.}
\end{figure}

In this section, we consider the interference channel discussed in the previous section, and describe our simulation results. For simplicity, we consider an interfernce channel with two 4-antenna transmitters and two single-antenna receivers, and plot the rate versus the logarithm of the transmit power. Note that we assume that the noise is zero-mean unit-variance Gaussian additive noise. 

The share-and-transmit scheme is implemented as explained in section \ref{subsection:SNT}. For the transmit only scheme, and since $T_1$ has a message for $R_1$ only, $T_1$ dedicates its available power to its link with $R_1$. Note that since the transmit signal space is much larger than the receive signal space, $T_1$ can decompose its channel with $R_1$ as well as its channel with $R_2$ to create one non-interfering link to $R_1$ and another to $R_2$. $T_2$ is able to achieve this as well, and dedicates all its power to its link with $R_2$. Each receiver can then decode its message without interference from other users.

In figure 3, we fix the distance between each transmitter and receiver to be equal to the distance between $T_1$ and $T_2$. In this case, the transmitters allocate the same resources to their sharing link as to their transmission links to the receivers. Figure 3 indicates that the share-and-transmit scheme always has a lower rate for the same transmit power than the original (transmit only) scheme. This result agrees with our analysis in section \ref{section:COOP}. 

In figure 4, we plot the result for the same interference channel as in figure 3, however when the distance between each transmitter and receiver is 5 times that between $T_1$ and $T_2$. Note that in this case, the sharing link is stronger than the transmission links since it does not suffer any path loss, whereas the transmission links do. Figure 4 shows that share-and-transmit scheme outperforms the original (transmit only) scheme. As expected, when the sharing link is stronger, cooperation between transmit nodes results in performance improvement over the non-cooperative scheme. Note that while our simulations are for the interference channel, similar results have been obtained for the multiple access scenario in \cite{Cui_Goldsmith_Bahai}.

\subsection{Remarks on X, Z and Relay Channels}
The X-channel is a general interference channel where both transmitters have messages for both receivers. Extending the proof for the interference channel, the (achievable) number of degrees of freedom for the X-channel can be shown to be that of the most capable MAC or BC channel. In other words,
\(\eta(X)\geq \max[\min(\max(M_1,M_2),N_1+N_2),\min(M_1+M_2,\max(N_1,N_2))]
\). Unlike the interference channel, non-trivial (other than PTP) outerbounds for the X-channel are not known even for single antennas at all nodes.  In particular, the results of \cite{Madsen_CTW} do not apply directly to the $X$ channel and no converse (or genie based outerbound) is known for the degrees of freedom even for the $(1,1,1,1)$ case.

The Z-channel is similar to the X-channel and the interference channel with the special property that one of the interfering channels is all zero. In other words, while there is a path between T2 and R1,  there is no path from T1 to R2. Notice that this makes the Z-channel similar to the channel obtained in the genie based outerbound when T1's message is provided by the genie to R2. Therefore, the corresponding results on the number of degrees of freedom for the Z-channel follow directly from the interference channel results presented in this paper. 

Finally, degrees of freedom can be computed in other distributed multiuser MIMO scenarios as well, such as the $(M_S,M_R,M_D)$ MIMO relay channel with $M_S$, $M_R$ and $M_D$ antennas at the transmitter, relay and destination respectively. Using the MAC and BC min-cut max-flow bounds on the relay channel it is easy to see that the number of degrees of freedom can not be increased by the presence of a relay, i.e.
\(\eta(RELAY)\leq\min[\min(M_S,M_R+M_D),\min(M_S+M_R,M_D)]=\min(M_S,M_D)\)
which is achievable through point to point communication between the source and the destination without a relay.
\section{Conclusions}\label{section:conclusion}

We investigate the available degrees of freedom for various multiuser MIMO communication scenarios such as the multiple access, broadcast, interference, relay, $X$ and $Z$ channels. Zero forcing is found to be optimal for achieving the degrees of freedom in all cases where the exact number of degrees of freedom is known. The distributed nature of the antennas significantly limits the degrees of freedom. For an interference channel with a total of $N$ transmit antennas and a total of $N$ receive antennas the available number of degrees of freedom can vary from $N$ to $1$ based on how the antennas are distributed among the two transmitters and receivers. Through an example of a share-and-transmit scheme, we show how the gains of transmitter cooperation are entirely offset by the cost of enabling that cooperation so that the available degrees of freedom are not increased.
\bibliographystyle{ieeetr}
\bibliography{Thesis}

\end{document}